\newcommand{\xba}{\alpha}
\newcommand{\xbb}{\beta}

\newcommand{\xbe}{\in}
\newcommand{\xbf}{\phi}

\newcommand{\xbm}{\mu}

\newcommand{\xbo}{\omega}

\newcommand{\xbq}{\psi}

\newcommand{\xbt}{\tau}

\newcommand{\xCN}{\neg}

\newcommand{\xCQ}{\emptyset}

\newcommand{\xCc}{<}

\newcommand{\xCf}{\hspace{0.1em}}

\newcommand{\xcB}{\subsetneqq}
\newcommand{\xcC}{\not\subseteq}

\newcommand{\xcE}{\exists}

\newcommand{\xcN}{\hspace{0.2em}\not\sim\hspace{-0.9em}\mid\hspace{0.8em}}

\newcommand{\xcP}{\not\rightarrow}

\newcommand{\xcc}{\subseteq}

\newcommand{\xce}{\not\in}

\newcommand{\xcg}{\geq}
\newcommand{\xch}{\Rightarrow}
\newcommand{\xci}{\Leftarrow}
\newcommand{\xcj}{\Leftrightarrow}

\newcommand{\xcn}{\hspace{0.2em}\sim\hspace{-0.9em}\mid\hspace{0.58em}}

\newcommand{\xco}{\vee}
\newcommand{\xcp}{\rightarrow}

\newcommand{\xcs}{\cap}

\newcommand{\xcu}{\wedge}
\newcommand{\xcv}{\cup}

\newcommand{\xcz}{\Box}

\newcommand{\xDH}{\item }

\newcommand{\xdN}{\mbox{\boldmath$N$}}

\newcommand{\xdf}{{\cal F}}

\newcommand{\xdi}{{\cal I}}

\newcommand{\xdm}{{\cal M}}

\newcommand{\xdp}{{\cal P}}

\newcommand{\xEI}{\begin{itemize}}
\newcommand{\xEJ}{\end{itemize}}

\newcommand{\xEc}{\not<}
\newcommand{\xEd}{\neq}

\newcommand{\xEh}{\begin{enumerate}}

\newcommand{\xEj}{\end{enumerate}}

\newcommand{\xEn}{\begin{description}}
\newcommand{\xEp}{\end{description}}

\newcommand{\xeb}{\prec}

\newcommand{\xee}{\succ}

\newcommand{\xfA}{\mid}

\newcommand{\Xl}{\ldots}

\newcommand{\bl}{\begin{lemma} \rm}
\newcommand{\el}{\end{lemma}}
\newcommand{\br}{\begin{remark} \rm}
\newcommand{\er}{\end{remark}}
\newcommand{\be}{\begin{example} \rm}
\newcommand{\ee}{\end{example}}
\newcommand{\bco}{\begin{corollary} \rm}
\newcommand{\eco}{\end{corollary}}
\newcommand{\bc}{\begin{claim} \rm}
\newcommand{\ec}{\end{claim}}
\newcommand{\bfa}{\begin{fact} \rm}
\newcommand{\efa}{\end{fact}}
\newcommand{\bp}{\begin{proposition} \rm}
\newcommand{\ep}{\end{proposition}}
\newcommand{\bd}{\begin{definition} \rm}
\newcommand{\ed}{\end{definition}}
\newcommand{\bcs}{\begin{construction} \rm}
\newcommand{\ecs}{\end{construction}}
\newcommand{\bcd}{\begin{condition} \rm}
\newcommand{\ecd}{\end{condition}}
\newcommand{\bt}{\begin{theorem} \rm}
\newcommand{\et}{\end{theorem}}
\newcommand{\bn}{\begin{notation} \rm}
\newcommand{\en}{\end{notation}}
\newcommand{\bfi}{\begin{bild} \rm}
\newcommand{\efi}{\end{bild}}
\newcommand{\bsta}{\begin{statement} \rm}
\newcommand{\esta}{\end{statement}}
\newcommand{\bcom}{\begin{comment} \rm}
\newcommand{\ecom}{\end{comment}}
\newcommand{\bdia}{\begin{diagram} \rm}
\newcommand{\edia}{\end{diagram}}

\newcommand{\bfc}{\begin{figure}[htb] \begin{center}}
\newcommand{\efc}{\end{center} \end{figure}}

\documentclass{article}

\usepackage{amssymb,latexsym,epic,eepic,rotating}

\title{Homogenousness and Specificity
\thanks{File: Hom, [Sch19a], arXiv 1902.09214
}
}

\author{Karl Schlechta
\thanks{
schcsg@gmail.com - https://sites.google.com/site/schlechtakarl/ -
Koppeweg 24, D-97833 Frammersbach, Germany}
\thanks{
Retired, formerly: Aix-Marseille Universit\'{e}, CNRS, LIF UMR 7279, F-13000
Marseille, France
}
}

\begin{document}

\newtheorem{lemma}{Lemma}[section]
\newtheorem{theorem}[lemma]{Theorem}
\newtheorem{proposition}[lemma]{Proposition}
\newtheorem{corollary}[lemma]{Corollary}
\newtheorem{claim}[lemma]{Claim}
\newtheorem{fact}[lemma]{Fact}
\newtheorem{remark}[lemma]{Remark}
\newtheorem{definition}{Definition}[section]
\newtheorem{construction}{Construction}[section]
\newtheorem{condition}{Condition}[section]
\newtheorem{example}{Example}[section]
\newtheorem{notation}{Notation}[section]
\newtheorem{bild}{Figure}[section]
\newtheorem{comment}{Comment}[section]
\newtheorem{statement}{Statement}[section]
\newtheorem{diagram}{Diagram}[section]

\renewcommand{\labelenumi}
  {(\arabic{enumi})}
\renewcommand{\labelenumii}
  {(\arabic{enumi}.\arabic{enumii})}
\renewcommand{\labelenumiii}
  {(\arabic{enumi}.\arabic{enumii}.\arabic{enumiii})}
\renewcommand{\labelenumiv}
  {(\arabic{enumi}.\arabic{enumii}.\arabic{enumiii}.\arabic{enumiv})}

\maketitle

\setcounter{secnumdepth}{3}
\setcounter{tocdepth}{3}

\begin{abstract}

We interpret homogenousness as a second order property and base it on
the same principle as nonmonotonic logic: there might be a small set of
exceptions.
We use this idea to analyse fundamental questions about defeasible inheritance
systems.

In an appendix, we discuss the concept of the core of a (model) set.

\end{abstract}

\tableofcontents
\clearpage

%
%
%
\section{
Introduction
}
\subsection{
Homogenousness as a default meta-rule
}

Homogeneousness was discussed as an important - though rarely explicitly
addressed - concept by the author in
 \cite{Sch97-2}, section 1.3.11, page 32,
and treated in more detail in
 \cite{GS16}, chapter 11, see also
 \cite{Sch18b}, section 5.7.

It is a second order hypothesis about the world, more precisely about
the adequacy of our concepts analysing the world, and discussed
in an informal way in
 \cite{GS16} and  \cite{Sch18b}.

The aim of these notes is to make the discussion more formal,
treating it as a second order application of the fundamental
concept of nonmonotonicity - that the set of exceptions is small -
and, in particular, to base the intuitively very appealing idea
of specificity - a way of solving conflicts between contradictory
homogenousness requirements -
on that same fundamental concept.

The author recently discovered
(reading  \cite{SEP13}, section 4.3)
that J. M. Keynes's
Principle of the Limitation of Independent Variety,
see  \cite{Key21} expresses essentially the same idea
as homogenousness.
(It seems, however, that the epistemological aspect, the naturalness
of our concepts, is missing in his work.)
By the way,  \cite{SEP13} also mentions
``inference pressure'' (in section 3.5.1)
discussed in
 \cite{Sch97-2}, section 1.3.4, page 10.
Thus, the ideas are quite interwoven.

Our main formal contribution here is to analyse a size relation
$<$ (or $<')$ between sets, generated by a relation $ \xeb $ between
elements - similarly to Definition 2.6 and Fact 2.7 in
 \cite{Sch97-2}.

We use these ideas to take a new look at defeasible inheritance systems
in Section 
\ref{Section Inher} (page 
\pageref{Section Inher}), and analyse two fundamental decisions
 \xEh
 \xDH Upward vs. downward chaining
 \xDH Extensions vs. direct scepticism
 \xEj
Moreoever we outline principles for a formal semantics based on our ideas.
\subsection{
A general comment
}

The reader will see that we treat here again semantics based on the
notions of distance and size. These notions seem very natural,
perhaps also because they have a neurological correspondence:
semantically close neurons or groups of neurons tend to fire
together, and a large number of neurons has a potentially bigger
effect than a small number, as their effect on other neurons might
add up.

In an appendix, we discuss a different, we think important, concept, the
core of
a set, base it on distance, and find it by repeated application of
standard
theory revision.
\section{
Filters and Ideals
}

\bd

$\hspace{0.01em}$


\label{Definition Filter}

Let $X \xEd \xCQ.$
 \xEh
 \xDH
$ \xdf (X) \xcc \xdp (X)$ is called a filter on $X$ iff

(1) $X \xbe \xdf (X),$ $ \xCQ \xce \xdf (X)$

(2) $A \xcc B \xcc X,$ $A \xbe \xdf (X)$ $ \xch $ $B \xbe \xdf (X)$

(3) $A,B \xbe \xdf (X)$ $ \xch $ $A \xcs B \xbe \xdf (X)$ (finite
intersection suffices here)

 \xDH
If there is $A \xcc X$ such that $ \xdf (X)=\{A' \xcc X:$ $A \xcc A' \},$
we say that $ \xdf (X)$
is the (principal) filter generated by $ \xCf A.$

 \xDH
$ \xdi (X) \xcc \xdp (X)$ is called an ideal on $X$ iff

(1) $X \xce \xdi (X),$ $ \xCQ \xbe \xdi (X)$

(2) $A \xcc B \xcc X,$ $B \xbe \xdi (X)$ $ \xch $ $A \xbe \xdi (X)$

(3) $A,B \xbe \xdi (X)$ $ \xch $ $A \xcv B \xbe \xdi (X)$ (finite union
suffices here)

 \xEj

\ed

\bd

$\hspace{0.01em}$


\label{Definition Correspondence}

Let $X \xEd \xCQ,$ $ \xdf (X)$ a filter over $X,$ then

$\{A \xcc X:$ $X-A \xbe \xdf (X)\}$ is the corresponding ideal $ \xdi (X)$
(and
$ \xdf (X) \xcs \xdi (X)= \xCQ).$

Given $ \xdf (X)$ and the corresponding $ \xdi (X),$ we set

$ \xdm (X):=\{A \xcc X:$ $A \xce \xdf (X) \xcv \xdi (X)\}.$

The intuition is that elements of the filter are big subsets, of the
ideal small subsets, and subsets in $ \xdm $ have medium size.

$ \xcz $
\\[3ex]

\ed

When speaking about $ \xdf,$ $ \xdi,$ $ \xdm $ over the same set $X,$ we
will always
assume that they correspond to each other as just defined.

\br

$\hspace{0.01em}$


\label{Remark <}

$X \xbe \xdi (X \xcv Y)$ $ \xch $ $Y \xbe \xdf (X \xcv Y),$ but not
necessarily the converse.

\er

\subparagraph{
Proof
}

$\hspace{0.01em}$


$X \xbe \xdi (X \xcv Y)$ $ \xch $ $(X \xcv Y)-X \xbe \xdf (X \xcv Y),$ and
$(X \xcv Y)-X \xcc Y,$ so $Y \xbe \xdf (X \xcv Y).$

For the converse: Consider $X=Y,$ then $Y \xbe \xdf (X \xcv Y),$ but $X
\xce \xdi (X \xcv Y).$

$ \xcz $
\\[3ex]

\bd

$\hspace{0.01em}$


\label{Definition <}

Given $X,$ $ \xdf (X)$ (and corresponding $ \xdi (X),$ $ \xdm (X)),$ and
$A,B \xcc X,$ we define:

 \xEh
 \xDH
$A<_{X}B$ $: \xcj $ $A \xbe \xdi (X),$ $B \xbe \xdf (X)$
 \xDH
$A<_{X}' B$ $: \xcj $

(a) $B \xbe \xdf (X)$ and $A \xbe \xdi (X) \xcv \xdm (X)$

or

(b) $B \xbe \xdm (X)$ and $A \xbe \xdi (X)$

 \xDH
If $X=A \xcv B,$ we write $A<B$ and $A<' B,$ instead of $A<_{X}B$ and
$A<_{X}' B.$

Note that case $(2)(b)$ of the definition is impossible if $X=A \xcv B:$
By Remark 
\ref{Remark <} (page 
\pageref{Remark <}), if $A \xbe \xdi (A \xcv B),$ then
$B \xbe \xdf (A \xcv B).$

 \xEj

Obviously, $<_{X}$ and $<_{X}' $ are irreflexive.

\ed

We define two coherence properties:

\bd

$\hspace{0.01em}$


\label{Definition Coherence}

(Coh1) $X \xcc Y$ $ \xch $ $ \xdi (X) \xcc \xdi (Y).$

(Coh2) $A,B \xbe \xdi (X),$ $A \xcs B= \xCQ $ $ \xch $ $A \xbe \xdi (X-
\xCf B)$

\ed

These properties will be discussed in more detail
below in Section 
\ref{Section Connections} (page 
\pageref{Section Connections}),
as they are closely
related to properties of a preferential relation
$ \xeb $ between elements of $X,$
see  \cite{Sch18a}.

First, an initial remark: if (Coh1) and (Coh2) hold, $ \xdf $ and $ \xdi $
behave well:

\bfa

$\hspace{0.01em}$


\label{Fact Subset}

$(Coh1)+(Coh2)$ imply:

(1)
Let $X \xbe \xdf (X'),$ then $(X \xcs A \xbe \xdf (X)$ $ \xcj $ $X' \xcs
A \xbe \xdf (X'))$

(2)
Let $X' \xbe \xdf (X),$ $Y' \xbe \xdf (Y),$ then the following four
conditions
are equivalent:

$X<Y,$ $X' <Y,$ $X<Y',$ $X' <Y' $

\efa

\subparagraph{
Proof
}

$\hspace{0.01em}$


(1)

``$ \xch $'': $X \xbe \xdf (X'),$ so $X' -X \xbe \xdi (X'),$ $X-A \xbe
\xdi (X) \xcc \xdi (X')$ by (Coh1), so
$(X' -X) \xcv (X-A) \xbe \xdi (X')$ $ \xch $ $X' -((X' -X) \xcv (X-A))=X'
\xcs X \xcs A \xcc X' \xcs A \xbe \xdf (X').$

``$ \xci $'': $X' \xcs A \xbe \xdf (X'),$ so $X' -A \xbe \xdi (X'),$
and $X-A \xcc X' - \xCf A,$ so $X-A \xbe \xdi (X').$
Moreover $X' -X \xbe \xdi (X'),$ and $(X' -X) \xcs (X-A)= \xCQ,$ so $X-A
\xbe \xdi (X)$ by (Coh2).

(2)

By Remark 
\ref{Remark <} (page 
\pageref{Remark <}), it suffices to show $X' \xbe \xdi (X' \xcv
Y)$ etc.

We use the finite union and downward closure properties of $ \xdi $
without
mentioning.

We also use the following without further mentioning:

(a) $(X \xcv Y)-(X-X') \xcc X' \xcv Y$

(b) $(X \xcv Y)-(Y-Y') \xcc X \xcv Y' $

(c) $(X \xcv Y)-((X-X') \xcv (Y-Y')) \xcc X' \xcv Y' $

(d) $X-X',$ $Y-Y',$ $(X-X') \xcv (Y-Y') \xbe \xdi (X \xcv Y)$ by
(Coh1)

We now show the equivalences.

(2.1) $X<Y$ $ \xch $ $X' <Y$:

$X' \xbe \xdi (X \xcv Y),$ $X-X' \xbe \xdi (X \xcv Y),$ so by (Coh2)
$X' $ $ \xbe $ $ \xdi ((X \xcv Y)-(X-X'))$ $ \xcc $ $ \xdi (X' \xcv Y)$
by (Coh1).

(2.2) $X<Y$ $ \xch $ $X<Y' $:

$X \xbe \xdi (X \xcv Y),$ $((Y-Y')- \xCf X)$ $ \xbe $ $ \xdi (Y)$ $ \xcc
$ $ \xdi (X \xcv Y)$ by (Coh1).
$X \xcs ((Y-Y')- \xCf X)$ $=$ $ \xCQ,$ so
$X \xbe \xdi ((X \xcv Y)-((Y-Y')- \xCf X))$ by (Coh2), but $(X \xcv
Y)-((Y-Y')- \xCf X)$ $=$ $X \xcv Y'.$

Note that we did not use $X',$ $X$ is just an arbitrary set.

(2.3) $X<Y$ $ \xch $ $X' <Y' $:

Let $X<Y,$ by (2.1) $X' <Y,$ so by (2.2) $X' <Y'.$

(2.4) $X<Y' $ $ \xch $ $X<Y$:

Trivial by (Coh1).

(2.5) $X' <Y$ $ \xch $ $X<Y$:

$X' \xbe \xdi (X' \xcv Y)$ $ \xcc $ $ \xdi (X \xcv Y),$ $X-X' \xbe \xdi
(X) \xcc \xdi (X \xcv Y),$ so
$X \xbe \xdi (X \xcv Y).$

Note that we did not use $Y',$ $Y$ is just an arbitrary set.

(2.6) $X' <Y' $ $ \xch $ $X<Y$:

Let $X' <Y',$ so $X' <Y$ by (2.4), so $X<Y$ by (2.5).

$ \xcz $
\\[3ex]

\bd

$\hspace{0.01em}$


\label{Definition Preference}

Let $X \xEd \xCQ,$ $ \xeb $ a binary relation on $X,$
we define for $ \xCQ \xEd A \xcc X$

$ \xbm (A)$ $:=$ $\{x \xbe A:$ $ \xCN \xcE x' \xbe A.x' \xeb x\}$

We assume in the sequel that for any such $X$ and $ \xCf A,$
$ \xbm (A) \xEd \xCQ.$

\ed

\bfa

$\hspace{0.01em}$


\label{Fact Mu}

Let $ \xdf (A):=\{A' \xcc A:$ $ \xbm (A) \xcc A' \}$ the filter over $
\xCf A$ generated
by $ \xbm (A),$ then the corresponding $ \xdi (A)=\{A' \xcc A:A' \xcs \xbm
(A)= \xCQ \},$
and $ \xdm (A)=\{A' \xcc A:$ $A' \xcs \xbm (A) \xEd \xCQ,$ and $ \xbm (A)
\xcC A' \}.$
$ \xcz $
\\[3ex]

\efa

When we discuss $ \xeb $ on $U,$ and $<_{X},$ $<,$ $<_{X}',$ $<' $ for
subsets of $U,$ we
implicitly mean the filters, ideals, etc. generated by $ \xbm $ on subsets
of $U,$
as discussed in Fact \ref{Fact Mu} (page \pageref{Fact Mu}).

It is now easy to give examples:

\be

$\hspace{0.01em}$


\label{Example Absolute}

$<$ is neither upward nor downward absolute.
Intuitively, in a bigger set, formerly big sets might become small,
conversely, in a smaller set, formerly small sets might become big.

Let $A,B \xcc X \xcc Y.$ Then

(1) $A<_{X}B$ does not imply $A<_{Y}B$

(2) $A<_{Y}B$ does not imply $A<_{X}B$

(1): Let $Y:=\{a,b,c\},$ $X:=\{a,b\},$ $c \xeb b \xeb a.$ Then
$\{a\}<_{X}\{b\},$ but
both $\{a\},\{b\} \xbe \xdi (Y).$

(2): Let $Y:=\{a,b,c\},$ $X:=\{a,c\},$ $c \xeb b \xeb a,$ but NOT $c \xeb
a.$
Then $\{a\}<_{Y}\{c\},$ but both $\{a\},\{c\} \xbe \xdm (X).$

\ee

We discuss now properties of $<$ and $<',$ and their relation to
properties of $ \xeb,$ when $<$ $(<')$ are generated by $ \xeb $ as in
Fact \ref{Fact Mu} (page \pageref{Fact Mu}).
\section{
$\xeb$ on U and $\xCc$ $(\xCc')$ on $\xdp(U)$
}

\label{Section Connections}

\bd

$\hspace{0.01em}$


\label{Definition Smooth}

We define the following standard properties for $ \xeb:$

(1) Transitivity (trivial)

(2) Smoothness

If $x \xbe X- \xbm (X),$ there there is $x' \xbe \xbm (X).x' \xeb x$

(3) Rankedness

If neither $x \xeb x' $ nor $x' \xeb x,$ and $x \xeb y$ $(y \xeb x),$ then
also $x' \xeb y$ $(y \xeb x').$

(Rankedness implies transitivity.)

See, e.g. Chapter 1 in  \cite{Sch18a}.
\subsection{
Simple and smooth $\xeb$
}

\ed

Recall:

\bd

$\hspace{0.01em}$


\label{Definition Pref-Rules}

$(\xbm PR)$ $X \xcc Y$ $ \xch $ $ \xbm (Y) \xcs X \xcc \xbm (X)$

$(\xbm CUM)$ $ \xbm (X) \xcc Y \xcc X$ $ \xch $ $ \xbm (X)= \xbm (Y)$

Again, see, e.g. Chapter 1 in  \cite{Sch18a}.

\ed

\bfa

$\hspace{0.01em}$


\label{Fact Coher}

(1) (Coh1) is equivalent to the basic property of preferential
structures, $(\xbm PR).$

(2) The basic property of smooth preferential structures, $(\xbm Cum),$
implies (Coh2), and $(Coh1)+(Coh2)$ imply $(\xbm Cum).$

\efa

\subparagraph{
Proof
}

$\hspace{0.01em}$


As there is a biggest $A \xbe \xdi (X),$ $A=X- \xbm (X),$ we can argue
with elements.

(1) $(\xbm PR)$ $ \xch $ (Coh1):
$x \xbe \xdi (X)$ $ \xch $ $x \xbe X,$ $x \xce \xbm (X)$ $ \xch $ $x \xbe
Y,$ $x \xce \xbm (Y).$

$(Coh1) \xch (\xbm PR):$
$x \xbe \xbm (Y) \xcs X,$ suppose $x \xce \xbm (X)$ $ \xch $ $x \xbe \xdi
(X)$ $ \xcc $ $ \xdi (Y)$ $ \xch $ $x \xbe Y,$ $x \xce \xbm (Y),$
contradiction.

(2) $(\xbm CUM)$ $ \xch $ (Coh2):
Let $A,B \xbe \xdi (X),$ $A \xcs B= \xCQ,$ so
$ \xbm (X-B)= \xbm (X)$ $ \xch $ $A \xbe \xdi (X- \xCf B).$

$(Coh1)+(Coh2)$ $ \xch $ $(\xbm CUM):$
Let $ \xbm (X) \xcc Y \xcc X.$
$X- \xCf Y,$ $Y- \xbm (X) \xbe \xdi (X),$ and $(X-Y) \xcs (Y- \xbm (X))=
\xCQ,$ so
$Y- \xbm (X) \xbe \xdi (X-(X-Y))= \xdi (Y),$ so $ \xbm (Y) \xcc \xbm (X).$
$ \xbm (X) \xcc \xbm (Y)$ follows from (Coh1)

$ \xcz $
\\[3ex]

\be

$\hspace{0.01em}$


\label{Example Non-Trans}

(1) Consider $a \xeb b \xeb c,$ but not $a \xeb c,$ with $Y:=\{b,c\},$
$X:=\{a,b\},$ $Z:=\{a,c\}.$
Then $\{b\}<_{X}\{a\},$ $\{c\}<_{Y}\{b\},$ but $\{c\} \xEc_{Z}\{a\}.$
Non-transitivity of $ \xeb $ is
crucial here.

(2) Consider $ \xCf a,$ $a_{i}:i \xbe \xbo,$ $b,$ $c$ with $c \xeb b,$ $a
\xee a_{0} \xee a_{1} \xee  \Xl.,$ and close under
transitivity. Then $\{a\}<_{\{a,b,a_{i}:i \xbe \xbo \}}\{b\},$
$\{b\}<_{\{b,c\}}\{c\},$ but $\{a\} \xEc_{\{a,c\}}\{c\}.$
Note that this structure is not smooth, but transitive.

$ \xcz $
\\[3ex]

\ee

\bfa

$\hspace{0.01em}$


\label{Fact <}

$<$ is transitive, if $ \xeb $ is smooth.

\efa

\subparagraph{
Proof
}

$\hspace{0.01em}$


By Fact 
\ref{Fact Coher} (page 
\pageref{Fact Coher}), we may use (Coh1) and (Coh2).

Let $X<Y<Z,$ so $X \xbe \xdi (X \xcv Y)$ and $Y \xbe \xdi (Y \xcv Z).$ We
have to show $X<_{X \xcv Z}Z,$ i.e.
$X \xbe \xdi (X \xcv Z),$ $Z \xbe \xdf (X \xcv Z).$

Consider $X \xcv Y \xcv Z,$ then by $X \xbe \xdi (X \xcv Y),$ $X \xbe \xdi
(X \xcv Y \xcv Z).$ By the same argument,
$Y \xbe \xdi (X \xcv Y \xcv Z),$ thus $Y-(X \xcv Z) \xbe \xdi (X \xcv Y
\xcv Z).$ As $(X \xcv Y \xcv Z)-(Y-(X \xcv Z))=X \xcv Z,$
and $X \xcs (Y-(X \xcv Z))= \xCQ,$ $X \xbe \xdi (X \xcv Z)$ by (Coh2),
and $Z \xbe \xdf (X \xcv Z)$ by
Remark \ref{Remark <} (page \pageref{Remark <})

$ \xcz $
\\[3ex]
\subsection{
Ranked $\xeb$
}

Rankedness speaks about $ \xdm,$ so it is not surprising that $<' $
behaves
well for ranked $ \xeb.$

\bd

$\hspace{0.01em}$


\label{Definition rk}

We define $rk(X):=rk(\xbm (X)).$

This is well-defined.

\ed

\bfa

$\hspace{0.01em}$


\label{Fact rk}

(1)

Let $A,B \xcc X.$

Then $A<_{X}' B$ iff

(a) $rk(B) \xeb rk(A)$ and $rk(B)=rk(X)$ or

(b) $rk(B)=rk(A)=rk(X)$ and $ \xbm (A) \xcB \xbm (B)= \xbm (X).$

(2)

$A<' B$ iff

(a) $rk(B) \xeb rk(A)$ or

(b) $rk(B)=rk(A)$ and $ \xbm (A) \xcB \xbm (B).$

(Recall that case (2) (b) in
Definition 
\ref{Definition <} (page 
\pageref{Definition <})  is impossible, if $X=A \xcv B).$

\efa

\subparagraph{
Proof
}

$\hspace{0.01em}$


(1)

$A<_{X}' B$ iff

$B \xbe \xdf (X)$ and $A \xbe \xdi (X)$ or

$B \xbe \xdf (X)$ and $A \xbe \xdm (X)$ or

$B \xbe \xdm (X)$ and $A \xbe \xdi (X).$

(2)

The case $A<' B$ is immediate.

$ \xcz $
\\[3ex]

\be

$\hspace{0.01em}$


\label{Example Trans-No-Rank}

Here, $ \xeb $ is transitive and smooth, but not ranked, and $<' $ is not
transitive.

\ee

Consider
$X:=\{x_{2},x_{3},x_{4}\},$ $Y:=\{x_{1},x_{2},y\},$

$x_{4} \xeb x_{2},$ $y \xeb x_{3},$ $y \xeb x_{1}.$

$ \xbm (X)=\{x_{3},x_{4}\},$ $\{x_{3}\} \xbe \xdm (X),$ $\{x_{2}\} \xbe
\xdi (X),$ $\{x_{2}\}<_{X}' \{x_{3}\}.$

$ \xbm (Y)=\{x_{2},y\},$ $\{x_{2}\} \xbe \xdm (Y),$ $\{x_{1}\} \xbe \xdi
(Y),$ $\{x_{1}\}<_{Y}' \{x_{2}\}.$

Let $x_{1},x_{3} \xbe Z,$ $\{x_{1}\}<_{Z}' \{x_{3}\}?$

If $y \xbe Z,$ $\{x_{1}\},\{x_{3}\} \xbe \xdi (Z).$

If $y \xce Z,$ $\{x_{1}\},\{x_{3}\} \xbe \xdm (Z).$

So $\{x_{1}\},\{x_{3}\}$ have same size in $Z.$
$ \xcz $
\\[3ex]

\bfa

$\hspace{0.01em}$


\label{Fact Trans-Rank}

Let the relation $ \xeb $ be ranked. Then $<' $ is transitive.

\efa

\subparagraph{
Proof
}

$\hspace{0.01em}$


Let $A<' B<' C.$
If both $A<' B$ and $B<' C$ hold by case (2) (b) in
Fact 
\ref{Fact rk} (page 
\pageref{Fact rk}), then $A<' C$ again by case (2) (b),
otherwise $A<' C$ by case (2) (a).

$ \xcz $
\\[3ex]

\bfa

$\hspace{0.01em}$


\label{Fact < m}

Let $X<Y$ $Y' \xbe \xdm (Y),$ then $X<Y' $ (and $X' <Y' $ for $X' \xbe
\xdm (X) \xcv \xdi (X)).$

(This does not hold for $X<' Y,$ of course.)

\efa

\subparagraph{
Proof
}

$\hspace{0.01em}$


By $X<Y,$ $rk(Y) \xeb rk(X),$ but $rk(Y)=rk(Y'),$ so $X<Y'.$
$ \xcz $
\\[3ex]
\section{
Specificity and Differentiation of Size
}

We now base the specificity criterion on the same notion of size
as nonmonotonicity.

In this section, $ \xcp $ and $ \xcP $ are the positive or negative arrows
of defeasible
inheritance diagrams.

\bfa

$\hspace{0.01em}$


\label{Fact Triangle}

Suppose (Coh1) holds.
If $C \xcc B$ or $C \xcp B,$ and $B \xcp A,$ $C \xcP A$ for some $ \xCf
A,$ then $C<B.$

(Likewise, if $B \xcP A,$ $C \xcp A,$ only the contradiction matters.)

\efa

\subparagraph{
Proof
}

$\hspace{0.01em}$


It suffices to show $C \xbe \xdi (B \xcv C),$ as then
$B \xbe \xdf (B \xcv C)$ by Remark \ref{Remark <} (page \pageref{Remark <}).

If $C \xcc B$ or $C \xcp B,$ then $C \xcs B \xbe \xdf (C),$ moreoever $C
\xcs \xCN A \xbe \xdf (C),$ so $C \xcs B \xcs \xCN A \xbe \xdf (C)$
by the finite intersection property.
Thus $C-(B \xcs \xCN A) \xbe \xdi (C) \xcc \xdi (B \xcv C)$ by (Coh1).
$C \xcs (B \xcs \xCN A) \xcc B \xcs \xCN A \xbe \xdi (B)$ by $B \xcp A,$
so
$C \xcs (B \xcs \xCN A) \xbe \xdi (B) \xcc I(B \xcv C)$ by (Coh1),
thus $C \xbe \xdi (B \xcv C)$ by the finite union property.

$ \xcz $
\\[3ex]

\bco

$\hspace{0.01em}$


\label{Corollary Triangle}

If $B \xcp A,$ $C \xcP A,$ then $C \xcp B$ and $B \xcp C$ together are
impossible by irreflexivity.

$ \xcz $
\\[3ex]

\eco

We may now base the specificity principle, like nonmonotonicity itself,
on small exception sets, but this time, the exceptions are second order.
We thus have a uniform background principle for reasoning.
\subsection{
Summary
}

It seems best to illustrate the situation with an example.

Consider the Tweety Diagram: $D \xcp B \xcp A,$ $D \xcp C \xcP A,$ $C \xcp
B.$
If we treat $D$ like $C,$ we violate a comparatively smaller subset:
As $C<B,$ and $D \xcp B,$ $D \xcp C,$ $D$ is a comparatively smaller
subset of $B$ than of $C,$
and smaller exception sets are more tolerable than bigger exception
sets.

Thus, size is a very strong concept for the foundation of
nonmonotonic reasoning.

(In addition, otherwise, the chain $D \xcp C \xcp B$ has two changes: $B
\xcp A,$ $C \xcP A,$ $D \xcp A,$
but this way, we have only one change: $B \xcp A,$ $C \xcP A,$ $D \xcP
A.)$
\subsection{
Differentiation of Size
}

Our basic approach has only three sizes: small, medium, big.

Using above ideas, we may further differentiate: if $A \xbe \xdi (B),$ and
$A' \xbe \xdi (A),$ then $A' $ is doubly small in $B,$ etc.
\section{
Homogenousness
}

We have now the following levels of reasoning:

 \xEh
 \xDH Classical logic:

monotony, no exceptions, clear semantics

 \xDH Preferential logic:

small sets of exceptions possible, clear semantics, strict rules about
exceptions, like $(\xbm CUM),$ no other restrictions

 \xDH Meta-Default rules (Homogenousness):

They have the form: $ \xba \xcn \xbb,$ and even if $ \xba \xcu \xba'
\xcN \xbb $ in the nonmonotonic
sense of (2), we prefer those models where $ \xba \xcu \xba' \xcn \xbb,$
but exceptions are
possible by nonmonotonicity itself, as, e.g., $ \xba \xcu \xba' \xcn \xCN
\xbb $ in (2).

We minimize those exceptions, and resolve conflicts whenever possible,
as in Fact \ref{Fact Triangle} (page \pageref{Fact Triangle}),
by using the same principle as in level (2): we keep
exception sets small. This is summarized in the specificity criterion.

(We might add a modified length of path criterion as follows:
Let $x_{0} \xcp x_{1} \xcp  \Xl  \xcp x_{n},$ $x_{i} \xcp y_{i},$ $x_{i+1}
\xcP y_{i}.$ We know by
Fact 
\ref{Fact Triangle} (page 
\pageref{Fact Triangle})  that $x_{0}< \Xl <x_{n},$ then any shorter
chain $a \xcp  \Xl  \xcp b$ has a
shorter possible size reduction (if there are no other chains, of
course!),
and we can work with this. This is the same concept as in
 \cite{Sch18e}, section 4.)

This has again a clear (preferential) semantics, as our characterisations
are abstract, see e.g.  \cite{Sch18a}.

 \xEj

Remark:
Inheritance diagrams and Reiter defaults are based on homogenousness, e.g.
in
the concatenation by default.
\section{
Extensions
}

All distance based semantics, like theory revision, counterfactuals,
have a natural notion of size: the set of closest elements is the smallest
set in the filter. Thus, we can apply our techniques to them, too.

Analogical reasoning and induction also operate with distance (analogical
reasoning) and size (comparison of the sample size with the target size),
so we may apply our principles here, too.
\clearpage
\section{
Defeasible Inheritance Diagrams
}

\label{Section Inher}

We discuss two of the main questions about defeasible
inheritance diagrams in the light of our above analysis.

 \xEh
 \xDH
Upward versus downward chaining
 \xDH
Extension based versus directly sceptical approaches
 \xEj
\subsection{
Upward versus downward chaining
}

\vspace{10mm}

\begin{diagram}

\label{Diagram Up-Down-Chaining}

\unitlength1.0mm

\begin{picture}(130,100)

\put(0,95){{\rm\bf The problem of downward chaining}}

\put(43,27){\vector(1,1){24}}
\put(37,27){\vector(-1,1){24}}
\put(13,57){\vector(1,1){24}}
\put(67,57){\vector(-1,1){24}}

\put(53,67){\line(1,1){4}}

\put(67,54){\vector(-1,0){54}}

\put(40,7){\vector(0,1){14}}
\put(43,7){\line(3,5){24}}
\put(58,28.1){\line(-5,3){3.6}}

\put(39,3){$Z$}
\put(39,23){$U$}
\put(9,53){$V$}
\put(69,53){$X$}
\put(39,83){$Y$}

\end{picture}

\end{diagram}

\vspace{4mm}

Discussion of Diagram 
\ref{Diagram Up-Down-Chaining} (page 
\pageref{Diagram Up-Down-Chaining}):

We assume (Coh1) and (Coh2).

By Fact 
\ref{Fact Triangle} (page 
\pageref{Fact Triangle}), we know that $X<V,$ and
by Fact \ref{Fact Subset} (page \pageref{Fact Subset})  (2), we know that
$X' <V' $ for any $X' \xbe \xdf (X),$ $V' \xbe \xdf (V),$ etc.

By specificity, there is $U' \xbe \xdf (U)$ with $U' \xcc X,$ $U' \xcc V,$
$U' \xcc \xCN Y.$

Of course, only (a big subset of) $U \xcs X$ is affected by $X \xcP Y,$
as we do only downward reasoning, no analogical (sideways) reasoning, or
so.

There is $Z' \xbe \xdf (Z)$ with $Z' \xcc U,$ $Z' \xcc \xCN X,$ thus, $Z'
$ is not affected
by $X \xcP Y.$

But there is no information that $Z' \xcC V.$

So there is $Z'' \xcc Z',$ $Z'' \xbe \xdf (Z),$ $Z'' \xcc U \xcs V \xcs
\xCN X,$ and $Z'' $ inherits from $V$
that $Z'' \xcc Y$ (or, better: there is $Z''' \xcc Z'',$ $Z''' \xbe
\xdf (Z),$ $Z''' \xcc Y).$

So, in this example, our (downward) approach coincides with upward
chaining, see  \cite{Sch97-2}, section 6.1.3. Basically, the reason
is that we look
inside $U,$ at $U \xcs X,$ $U- \xCf X,$ and not only globally at $U,$
which would involve
(hidden) analogous reasoning.
\subsection{
Extension based versus directly sceptical approaches
}

Consider Diagram \ref{Diagram Nixon} (page \pageref{Diagram Nixon}).

Extension based approaches branch into different extensions when
a conflict cannot be solved by specificity.

Each extension violates the homogenousness assumption
rather drastically. Assuming that half of $U$ (a medium size subset) is in
$Y,$ the
other half in $ \xCN Y,$ is a less drastic violation, thus it corresponds
to
the overall strategy to minimize violation of homogenousness.

But, there is no prinpal difference between two medium size subsets
and one big and one small subset which are in conflict. So they
should be treated the same way. Thus, in one single picture we have not
only
conflicting big and small subsets, but also conflicting medium size
subsets, so this is more in the directly sceptical colour - without
saying it is strictly the same as the traditional directly sceptical
approach.

Thus, we have $U' \xbe \xdf (U)$ with $U' \xcc V \xcs X,$ and $U_{1}'
,U_{2}' \xbe \xdm (U'),$
$U_{1}' =U' -U_{2}',$ $U_{1}' \xcc Y,$ $U_{2}' \xcc \xCN Y.$

Of course, any $U'' \xcc U-X$ is NOT affected by $X \xcP Y.$

Thus, considering Diagram 
\ref{Diagram Extended-Nixon} (page 
\pageref{Diagram Extended-Nixon}),
there is $Z' \xbe \xdf (Z),$ $Z' \xcc U- \xCf X,$ and this inherits only
from $V,$ i.e.
that it is mostly in $Y.$

\vspace{10mm}

\begin{diagram}

\label{Diagram Nixon}

\unitlength1.0mm

\begin{picture}(130,100)

\put(0,95){{\rm\bf The Nixon Diamond}}

\put(43,27){\vector(1,1){24}}
\put(37,27){\vector(-1,1){24}}
\put(13,57){\vector(1,1){24}}
\put(67,57){\vector(-1,1){24}}

\put(53,67){\line(1,1){4}}



\put(39,23){$U$}
\put(9,53){$V$}
\put(69,53){$X$}
\put(39,83){$Y$}

\end{picture}

\end{diagram}

\vspace{4mm}

\vspace{10mm}

\begin{diagram}

\label{Diagram Extended-Nixon}

\unitlength1.0mm

\begin{picture}(130,100)

\put(0,95){{\rm\bf Extended Nixon Diamond}}

\put(43,27){\vector(1,1){24}}
\put(37,27){\vector(-1,1){24}}
\put(13,57){\vector(1,1){24}}
\put(67,57){\vector(-1,1){24}}

\put(53,67){\line(1,1){4}}


\put(40,7){\vector(0,1){14}}
\put(43,7){\line(3,5){24}}
\put(58,28.1){\line(-5,3){3.6}}

\put(39,3){$Z$}
\put(39,23){$U$}
\put(9,53){$V$}
\put(69,53){$X$}
\put(39,83){$Y$}

\end{picture}

\end{diagram}

\vspace{4mm}

\be

$\hspace{0.01em}$


\label{Example Two-Nixon}

Let $U$ be the bottom node of two Nixon diamonds, e.g. add
to Diagram 
\ref{Diagram Nixon} (page 
\pageref{Diagram Nixon})  three nodes $V',X',Y' $
with $U \xcp V' \xcp Y',$ $U \xcp X' \xcP Y'.$

Then we have to split $U$ into two sets $U_{0},U_{1} \xbe \xdm (U),$ with
$U_{0} \xcc Y,$ $U_{1} \xcs Y= \xCQ $
(basically) and two sets $U_{2},U_{3} \xbe \xdm (U),$ with $U_{2} \xcc Y'
,$ $U_{3} \xcs Y' = \xCQ,$
and, as they are independent, we split $U$ into four sets, all in $ \xdm
(U),$
$U_{00},$ $U_{01},$ $U_{10},$ $U_{11},$ with e.g. $U_{00} \xcc Y,Y',$
$U_{01} \xcc Y,$ but $U_{01} \xcs Y' = \xCQ,$
$U_{10} \xcs Y= \xCQ,$ but $U_{10} \xcc Y',$ and $U_{11} \xcs Y= \xCQ,$
$U_{11} \xcs Y' = \xCQ.$

Note that it unimportant in which order we treat the conflicts or if
we treat them simultanously - as should be the case.
\subsection{
Ideas for a Semantics
}

\ee

If we want to treat the Nixon Diamond, we have to consider $ \xdm (X).$
So far, we have not considered abstract coherence properties for
$ \xdm.$ We do this now.

\bd

$\hspace{0.01em}$


\label{Definition  m}

$(\xbm =)$
This is a condition for ranked preferential structures.

$X \xcc Y,$ $X \xcs \xbm (Y) \xEd \xCQ $ $ \xch $ $ \xbm (X)= \xbm (Y)
\xcs X.$

We re-write this:

$X \xbe \xdm (Y) \xcv \xdf (Y)$ $ \xch $ $ \xdf (X)=\{Y' \xcs X:Y' \xbe
\xdf (Y)\}$

(The case $X \xbe \xdf (Y)$ does not interest here very much.)

\ed

Illustration of (the main part of) $(\xbm =):$

Suppose we add to Diagram 
\ref{Diagram Nixon} (page 
\pageref{Diagram Nixon})  an arrow $U \xcp W,$ then
we know that
$U \xcs W \xbe \xdf (U),$ and $U \xcs Y \xbe \xdm (U),$ $U-Y \xbe \xdm
(U),$ so $U \xcs W \xcs Y \xbe \xdf (U \xcs Y),$
$U \xcs W-Y \xbe \xdf (U- \xCf Y)$ by $(\xbm =).$

If, in addition, we add $Z,$ and $Z \xcp U,$ and a negative arrow $Z \xcP
Y$
(not $Z \xcP X,$ as in Diagram 
\ref{Diagram Extended-Nixon} (page 
\pageref{Diagram Extended-Nixon})),
$Z$ is mostly in $U' \xbe \xdm (U)$ with $U' \xcs Y= \xCQ,$ and we may
still conclude
by the above that $Z$ inherits to be (mostly) in $W$ from $U \xcs W-Y \xbe
\xdf (U- \xCf Y).$
\subsubsection{
Principles
}

We work with very few background principles:

 \xEh
 \xDH
For many purposes, reasoning with abstract size seems the
adequate approach.
 \xDH
As always in nonmonotonic reasoning, small sets of (first-level)
exceptions are
possible, so we work with $ \xdf $ and $ \xdi,$ instead of the full or
empty set
(we used $ \xCQ $ above as an abbreviation).
 \xDH
The hard rules of the background logic and of the filter/ideal
properties tell us how to treat big/small/medium subsets on the first
level.
 \xDH
This is complemented by the homogenousness principle and conflict
resolution by specificity on the second level.
 \xDH
We treat all subsets the same way, not medium size sets differently by
branching into different possibilities.
 \xDH
Specificity is based on the same idea as nonmonotonicity itself:
we tolerate (better) small exception sets (than bigger ones).

 \xEj

Based on these principles, we proceed as follows:

 \xEh

 \xDH
We decide for a background logic, i.e. for coherence conditions.
(Coh1), (Coh2), $(\xbm =)$ seems a good choice.

 \xDH
We respect the coherence conditions, and inherit properties strictly
downward, not by analogy. Contradictions are either solved by specificity,
or we chose one half for property $ \xbf,$ the other for $ \xCN \xbf.$

Independence is respected as in
Example \ref{Example Two-Nixon} (page \pageref{Example Two-Nixon}).

It is important to chose the (sub)sets from which we inherit carefully,
there is no analogical reasoning here.
This was illustrated in above examples.

 \xEj
\clearpage
\section{
Appendix - the Core of a Set
}

The following remarks are only abstractly related to the main part
of these notes. The concept of a core is a derivative concept to the
notion of a distance, and the formal approach is based on theory revision,
see e.g.  \cite{LMS01}, or  \cite{Sch18b},
section 4.3.

We define the core of a set as the subset of those elements which are
``sufficiently'' far away from elements which are NOT in the set.
Thus, even if we move a bit, we still remain in the set.

This has interesting applications. E.g., in legal reasoning, a witness may
not be very sure about colour and make of a car, but if he errs in one
aspect, this may not be so important, as long as the other aspect is
correct. We may also use the idea for a differentiation of truth
values, where a theory may be ``more true'' in the core of its
models than in the periphery, etc.

In the following, we have a set $U,$ and a distance $d$ between elements
of $U.$
All sets $X,Y,$ etc. will be subsets of $U.$ $U$ will be finite, the
intuition
is that $U$ is the set of models of a propositional language.

\bd

$\hspace{0.01em}$


\label{Definition Depth}

Let $x \xbe X \xcc U.$

(1) $depth(x)$ $:=$ $min\{d(x,y):$ $y \xbe U-X\}$

(2) $depth(X)$ $:=$ $max\{depth(x):$ $x \xbe X\}$

\ed

\bd

$\hspace{0.01em}$


\label{Definition Core}

Fix some $m \xbe \xdN,$ the core will be relative to $m.$
One might write $Core_{m},$ but this is not important here,
where the discussion is conceptual.

Define

$core(X)$ $:=$ $\{x \xbe X:$ $depth(x) \xcg depth(X)/m\}$

(We might add some constant like 1/2 for $m=2,$ so singletons have a
non-empty core - but this is not important for the conceptual
discussion.)

\ed

It does not seem to be easy to describe the core operator
with rules e.g. about set union, intersection, etc.
It might be easier to work with pairs (X, $depth(X)),$ but
we did not pursue this.

We may, however, base the notion of core on repeated application of
the theory revision operator $*$ (for formulas) or $ \xfA $ (for sets) as
follows:

Given $X \xcc U$ (defined by some formula $ \xbf),$ and $Y:=U-X$ (defined
by $ \xCN \xbf),$
the outer elements of $X$ (those of depth 1) are $Y \xfA X$ $(M(\xCN \xbf
* \xbf)).$
The elements of depth 2 are $(Y \xcv (Y \xfA X)) \xfA (X-(Y \xfA X)),$
M($((\xCN \xbf) \xco (\xCN \xbf * \xbf))$ $*$ $(\xbf \xcu \xCN (
\xCN \xbf * \xbf))$) respectively, etc.

We make this formal.

\bfa

$\hspace{0.01em}$


\label{Fact Core}

 \xEh
 \xDH The set version

Consider $X_{0},$ we want to find its core.

Let $Y_{0}$ $:=$ $U-X_{0}$

Let $Z_{0}$ $:=$ $Y_{0} \xfA X_{0}$

Let $X_{1}$ $:=$ $X_{0}-Z_{0}$

Let $Y_{1}$ $:=$ $Y_{0} \xcv Z_{0}$

Continue $Z_{1}:=$ $Y_{1} \xfA X_{1}$
etc. until it becomes constant, say $Z_{n}=X_{n}$

Now we go back: $Core(X_{0})$ $:=$ $X_{n} \xcv  \Xl  \xcv X_{n/2}$

 \xDH The formula version

Consider $ \xbf_{0},$ we want to find its core.

Let $ \xbq_{0}$ $:=$ $ \xCN \xbf_{0}$

Let $ \xbt_{0}$ $:=$ $ \xbq_{0}* \xbf_{0}$

Let $ \xbf_{1}$ $:=$ $ \xbf_{0} \xcu \xCN \xbt_{0}$

Let $ \xbq_{1}$ $:=$ $ \xbq_{0} \xco \xbt_{0}$

Continue $ \xbt_{1}:=$ $ \xbq_{1}* \xbf_{1}$
etc. until it becomes constant, say $ \xbt_{n}= \xbf_{n}$

Now we go back: $Core(\xbf_{0})$ $:=$ $ \xbf_{n} \xco  \Xl  \xco
\xbf_{n/2}$

 \xEj
\subsection{
Acknowledgement
}

\efa

The author would like to thank David Makinson
for an important comment.

\end{document}